\documentclass{amsart}

\usepackage[pagewise]{lineno}\linenumbers
\usepackage[utf8]{inputenc}
\usepackage{amsthm,amsmath,amsfonts}
\usepackage{soul,color}
\usepackage{longtable}
\usepackage{xcolor}
\usepackage{tabularx}
\usepackage{longtable}
\usepackage{booktabs}
\usepackage{mathtools}
\usepackage{mathabx}
\newcommand{\Fq}{\mathbb{F}_q}
\newtheorem{theorem}{Theorem}[section]

\newtheorem{lemma}[theorem]{Lemma}

\usepackage{color,soul,xcolor}
\usepackage{multicol}
\usepackage{enumitem}
\usepackage{hyperref}
\usepackage[ruled,vlined]{algorithm2e}

\usepackage{comment}
\theoremstyle{definition}
\newtheorem{definition}{Definition}[section]

\title{New Binary and Ternary Quasi-Cyclic Codes with Good Properties}
\author{Dev Akre, Nuh Aydin, Matthew J. Harrington, Saurav R. Pandey}
\date{}

\begin{document}
\nolinenumbers

\maketitle
\begin{abstract}
One of the most important and challenging problems in coding theory is to construct codes with best possible parameters and properties. The class of quasi-cyclic (QC) codes is known to be fertile to produce such codes. Focusing on QC codes over the binary field, we have found 113 binary QC codes that are new among the class of QC codes using an implementation of a fast cyclic partitioning algorithm and the highly effective ASR algorithm. Moreover, these codes have the following additional properties: a) they have the same parameters as best known linear codes, and b) many of the have additional desired properties such as being reversible, LCD, self-orthogonal or dual-containing. Additionally, we present an algorithm for the generation of new codes from QC codes using ConstructionX, and introduce 35 new record breaking linear codes produced from this method.
\end{abstract}


\section{Introduction and Motivation}

\bigskip

A linear block code $C$ of length $n$ over the finite field $GF(q)$ (the code alphabet, also denoted by $\Fq$) is a vector subspace of $\Fq^n.$ If the dimension of $C$ is  $k$ and its minimum distance is $d$, then $C$ is referred to as an $[n,k,d]_q$-code. Elements of $C$ are called codewords. A matrix whose rows constitute a basis for $C$ is called a generator matrix of $C$.

One of the main goals of coding theory is to construct  codes with best possible parameters.  This is an optimization problem that  can be formulated in a few different ways. For example, we can fix $n$ and $k$ (hence the information rate of the code) and ask for the largest possible value $d_q[n,k]$ of $d$. A code of length $n$ and dimension $k$  whose minimum distance is $d_q[n,k]$ is an optimal code (or a distance-optimal code). Similarly, one can fix any two of the three parameters of a linear code and look for the optimal value of the third parameter. There are many theoretical bounds on the parameters of a linear code.  There are also databases of best known linear codes (BKLC). The online database \cite{database} is well known in the coding theory research community. For a given set of $q, n$, and $k$, the database gives information about\\
a) the best theoretical upper bound for $d$, and\\
b) the highest minimum distance of a best known linear code with the given length and the dimension which is a lower bound on $d$.\\
Lower bounds are usually obtained through explicit constructions. A code with an explicit construction that has the largest known minimum distance for these parameters provides a lower bound. It is possible to obtain codes with the same parameters in multiple ways and with different structures. In most cases, there are gaps between theoretical upper bounds and lower bounds. The Magma software \cite{magma} also has a similar database. Additionally, there are also more specialized databases such as the one specifically for quasi-cyclic  (QC) and quasi-twisted (QT) codes \cite{qcdatabase}.

%A special class of codes called linear codes are much more efficient to work with than an arbitrary set of codes. They require less storage space and are faster to encode than arbitrary non-linear codes.

Constructing codes with best possible parameters is a challenging problem. It  is clear from the databases that in most cases optimal codes are not yet known. They are usually known when either $k$ or $n-k$ is small. Hence, there are many instances of this optimization problem that are open. For example, for $q=5$ and $n=100$, there are gaps between the upper bounds and lower bounds on $d$ for every dimension  $5\leq k \leq 91$. 
%The approximate symmetry in this situation is due to the fact that the minimum distance of a linear code (dimension $k$) is related to the minimum distance of its dual (dimension $n-k$) via the well known MacWilliam's identity.    

There are two main reasons why this optimization problem is very challenging even with the help of modern computers. First, determining the minimum distance of a linear code is computationally intractable (\cite{NPhard}) so it takes significant amount of time to find the minimum distance of a single code when the dimension is large (and becomes infeasible after a certain point). Second, for a given length and dimension, the number  $\displaystyle{\frac{(q^n-1)(q^n-q)\cdots (q^n-q^{k-1})}{(q^k-1)(q^k-q)\cdots (q^k-q^{k-1})}}$ of linear codes of length $n$ and dimension $k$ over $\Fq$ is  large  and grows quickly. Hence, an exhaustive computer search on linear codes is not feasible. Therefore, researchers  focus on specific classes of codes with rich mathematical structures that are known to contain many codes with good parameters. The class of quasi-cyclic (QC) codes has an excellent record of producing many codes with best known parameters. They are the focus of our search in this work. In fact, we have combined three methods that are known to be useful in coding theory: a) computer searches over QC codes b) using a recently introduced algorithm to test equivalence of two cyclic codes c) Construction X. 

 Our search revealed 113  binary QC codes that are new among the class of QC codes according to \cite{qcdatabase}. Moreover, our codes also have the following additional features:
\begin{itemize}
    \item Each of these codes has the same parameters as BKLCs in \cite{database}.
    \item In many cases the BKLCs in the database (\cite{database}) have indirect, multi-step constructions so it is more efficient and desirable to obtain them in the form of QC codes instead.
    \item Many of our codes have additional desirable properties such as  being self-orthogonal, reversible, or linear complementary dual (LCD).
\end{itemize}

Additionally, we applied a ConstructionX method that uses QC codes with good parameters as the input and found 35 new linear codes of which 8 are binary.

The rest of the paper is organized as follows. In the next section we present some basic definitions that are fundamental to our work. In section 3, we explain our search method. Finally, we list the new codes in the last section.

%\newpage

\section{Basic Definitions}

%\bigskip

%\hl{Cyclic codes are a generalization of linear codes.} \hlc[pink]{No!}.\\
%\hl{Since we will send this to a different journal now, I will modify this section.}
Cyclic codes have a prominent place in coding theory for both theoretical and practical reasons.  Some of the best known examples of codes are eiter cyclic or equivalent to cyclic codes including binary Hamming codes, the Golay codes, BCH codes, Reed-Solomon codes, and quadratic residue codes to name a few. They are conveniently implemented via shift registers.  Theoretically, they establish a key link between coding theory and algebra.  The first step in this connection is to represent a vector $(c_0,c_1,\dots,c_{n-1})$ in $\Fq^n$ as the polynomial $c(x)=c_0+c_1x+\cdots+c_{n-1}x^{n-1}$ of degree less than $n$. This correspondence defines a vector space isomorphism between $\Fq^n$ and the set of polynomials of degree $< n$ over $\Fq$. With this identification, we use vectors/codewords and polynomials interchangeably. 

\begin{definition}
A linear code $C$ is called cyclic if it is closed under the cyclic shift $\pi$, i.e. whenever $c=(c_0,c_1,...,c_{n-1})$ is a codeword of C, then so is 
$\pi(c) = (c_{n-1},c_0,.....,c_{n-2})$. 
\end{definition}
%\medskip

In the polynomial representation, the cyclic shift of a codeword $c(x)$ corresponds to $xc(x) \mod x^n-1$. It follows that a cyclic code is an ideal in the quotient ring $\Fq[x]/\langle x^n-1 \rangle$ which is a principal ideal ring. Hence any cyclic code $C$ can be viewed as a principal ideal $C=\langle g(x)\rangle=\{ f(x)g(x) \mod x^n-1: f(x) \in \Fq[x]\}$ generated by $g(x)$. A cyclic code $C$ has many  generator polynomials and among them is a unique one. The monic, non-zero polynomial of least degree in $C$ is a unique generator for $C$. We will refer to this unique generator as the (standard) generator of $C$. When speak of ``the generator polynomial'' of a cyclic code, the standard generator should be understood.

The following are well known about cyclic codes.

\begin{lemma}
Let $C=\langle g(x)\rangle$ be a  cyclic code of length $n$ over $\Fq$ where $g(x)$  is the standard generator polynomial. Then the following holds

\begin{enumerate}
    \item $g(x)$ is a divisor of $x^n-1$ over $\Fq$. Hence $x^n-1=g(x)h(x)$ for some $h(x)\in \Fq[x]$.
    
    \item  The polynomial $h(x)$ is called the check polynomial and it has the property that  a word $v(x)$ is in $C$ if and only if $h(x)v(x)=0$ in $\mathbb{F}_q[x]/ \langle x^n-1 \rangle$.
    
   \item  The dimension of $C$ is $k=n-deg(g(x))=\deg(h(x))$ and  a basis for $C$ is $\{g(x),xg(x),...,x^{k-1}g(x)\}$.
   
   \item If $g(x)=g_0+g_1x+\cdots+g_{r}x^r$ then $g_0\not = 0$ and the following circulant matrix is a generator matrix for $C$, where each row is a cyclic shift of the previous row.
   
   \begin{displaymath}
G=\left[\begin{array}{cccccccccc}
g_0& g_1 & g_2  & \ldots  &g_r    &0  &0 &\ldots &0\\
0&   g_0 & g_1 &  g_2     &\ldots &g_r&0&\dots &0 \\
\vdots& \ddots  & \ddots  & \ddots  & \ddots & \ddots  & \ddots & &\vdots \\
0& 0& 0&\ldots &0 & g_0 & g_1&  \ldots &g_r
\end{array}
\right]
\end{displaymath}

%   \[
 %G=
%  \begin{bmatrix}
%    g_0 & g_{1} & g_{2} & \cdots & g_{r} & 0 \cdots 0 \\
%    0 & g_{0} & g_{1} & \cdots & g_{r-1} & g_r \cdots 0 \\
%    \vdots & \vdots &  \ddots & \ddots &  \ddots  & \vdots \\
%    0 & 0 & \cdots & g_0 & \cdots & g_{r}
%  \end{bmatrix}
%\]

   \item $C=\langle p(x) \rangle$ if and only if $p(x)=f(x)g(x)$ where $\gcd(f(x),h(x))=1$.
   
   \item There is a one-to-one correspondence between divisors of $x^n-1$ and cyclic codes of length $n$ over $\Fq$.
    
\end{enumerate}

\end{lemma}

%\medskip

In the study of cyclic codes, the notion of cyclotomic cosets is fundamental.  There is a one-to-one correspondence between cyclotomic cosets of $q \mod n$ and irreducible divisors of $x^n-1$ over $\Fq$. The important class of BCH codes are defined based on cyclotomic cosets. We will make use of cyclotomic cosets in our search process.    
%Let the positive integer $n$ be relatively prime with $q$. Define a relation $\sim$ on integers $\mathbb{Z}_n=\{0,1,\dots,n-1 \} \quad \mod n$  by $x\sim y$ if $x=q^iy$ for some integer $i$. This is an equivalence relation $\mathbb{Z}_n$.  Each equivalence class is of the form $C_a=\{ a, aq,aq^2,\dots,aq^{r-1} \}$ where $r$ is the smallest positive integer such that $aq^r\equiv a \mod n$ and it is called a cyclotomic coset, more specifically the cyclotomic coset of $q \mod n$ containing $a \in \mathbb{Z}_n $. 

Cyclic codes have many useful generalizations. One of the most important generalizations is quasi-cyclic (QC) codes where we can shift coordinates of codewords by more than one positions.
\newpage
%Quasi-cyclic (QC) codes are generalizations of cyclic codes where the shift can occur by $\ell$ positions, with a cyclic code being represented by the case $\ell = 1$. 

\begin{definition}
A linear code $C$ is said to be $\ell$-quasi-cyclic (QC) if for a positive integer $\ell$, whenever $c = (c_0,c_1,...,c_{n-1})$ is a codeword, so is \\ $(c_{n-\ell},c_{n-\ell+1},...,c_{n-1}, c_0, c_1,...,,c_{n-\ell-2},c_{n-\ell-1})$. Such a code $C$ is called a QC code of index $\ell$, or an $\ell$-QC code.
\end{definition}
%\medskip

It is well known that the length of a QC code must be a multiple of $\ell$, hence $n=m\ell$ for some positive integer $m$ \cite{ASR}.  Let $R_m=\frac{\Fq[x]}{\langle
x^m-1\rangle}$. Then an $\ell$-QC code is an $\Fq[x]$-submodule of $R_m^{\ell}$  \cite{ASR}. QC codes are known to contain many codes with good parameters. Hundreds of BKLCs in \cite{database} are obtained from QC codes. A particularly effective search algorithm called ASR was presented in \cite{ASR} and has been employed many times since then (e.g., \cite{daskalov2003}, \cite{ackerman2011}, \cite{gf7}, \cite{Twistulant} ). This is the basis of  the method we use in this work as well.

For any linear code $C$, its dual code is defined as $C^\perp=\{ v\in \Fq^n: v\cdot c=0 \text{ for all } c\in C \}$ where $v\cdot c$ is the standard inner product in $\Fq^n$. If the dimension of $C$ is $k$, then the dimension of $C^\perp$ is $n-k$.  A code $C$ is self-orthogonal if  $C \subseteq C^\perp$, i.e., for any two codewords $a, b \in C$, $a \cdot b = 0$. An $[n,k]_q$ code $C$ is self-dual if  $C = C^\perp$. Note that in this case, the dimensions of $C$ and $C^\perp$ need to be equal. Thus,  $k = n/2$. A code $C$ is dual-containing if $ C^\perp \subseteq C$.\\

Self dual codes are an important area of research and there is a vast literature about them in coding theory.  One application of self-orthogonal codes is in constructing  quantum error correcting codes (QECC) from classical codes. A method of constructing quantum error correcting codes (QECC) from classical codes was given in \cite{Quantumoriginal3}. Since then researchers have investigated various methods of using classical error correcting codes to construct new QECCs.
%Some researchers have studied constructing QECCs from additive constacyclic codes in the paper "New quantum codes from additive constacyclic codes" (currently under review) but 
The majority of the methods have been based on the CSS construction given  in \cite{Quantumoriginal3}. In this method, self-dual, self-orthogonal and dual-containing linear codes are used to construct quantum codes. The CSS construction requires two linear codes $C_1$ and $C_2$ such that $C_2^{\perp}\subseteq C_1$. Hence, if $C_1$ is a self-dual code, then we can construct a CSS quantum code using $C_1$ alone since $C_1^{\perp}\subseteq C_1$. If  $C_1$ is self-orthogonal, then we can construct a CSS quantum code with $C_1^{\perp}$ and $C_1$ since $C_1 \subseteq C_1^{\perp}$. Similarly in the case $C_1$ is a dual-containing code. In a recent work, many new best known quantum codes (BKQC) have been found from classical self-orthogonal codes \cite{mtgood}.
 
 %As $C \subseteq C^\perp$, there is a constraint on their dimensions,  $n < n-k$, which gives $k< n/2$.

\medskip

QC codes generated by the ASR algorithm \cite{ASR} use good cyclic codes as building blocks. For the codes generated by this algorithm, the length is typically much larger than the dimension.  Thus, in most cases we have $k < n/2$. Consequently, we can only  find self-orthogonal codes from this method which can still be used in constructing quantum codes.

%\medskip
A code $C$ is linear complementary dual (LCD) if  $C \cap C^\perp = \{ 0 \}$. They were first introduced by Massey \cite{LCD}, and were seen to have an optimal solution for a two-user binary adder channel as well as decoding algorithms that are less complex than that for general linear codes. They are also useful in cryptography by protecting the information managed by sensitive devices, particularly against fault invasive attacks and side-channel attacks (SCA) \cite{LCDuse}. We have been able to find a number of QC codes that are  LCD  in our search.

\medskip

Another useful property of a code is being reversible. A code $C$ is reversible if  for any codeword $(w_0,w_1,...,w_{n-2},w_{n-1}) \in C$ its reverse $(w_{n-1},w_{n-2},...,w_1,w_0)$ is also in $C$. Suppose we have a reversible code $C$ stored in some storage medium. Since the reverse of every codeword of $C$ is also a codeword, the stored data can be read from either end of the code, which could be advantageous if, for example, we are interested in only the information at one end of the code. If the decoder has to read the entire code before beginning the decoding process, then the code being reversible is not important. However, if the code can be decoded digit by digit, then the same decoding circuit can be used irrespective of the end of the code that is fed first \cite{mas}. 

%\newpage
\section{QC Search Method}

Our goal in this search was to  find binary QC codes with good parameters and good properties. We employed the generalized version of the ASR algorithm described in \cite{GenASR} as our search method. The first step in the search  process is to obtain all cyclic codes for all lengths of interest and partition them into equivalence classes based on code equivalence.  Although Magma software has a command to test equivalence of linear codes, a more efficient method that is specifically for cyclic codes has recently been introduced in \cite{cycliceq}. We implemented  this algorithm in our work. The rest of this section gives more details on the search process. 

%original goal was to psuedo-exhaustively generate all QC codes over $\mathbb{F}_2$. To do this, we used the ASR search algorithm for QC code generation first described in \cite{ASR}.
%\bigskip

\subsection{Cyclic Partition Algorithm\\ \\}

%The first step to the ASR algorithm for QC codes is to generate cyclic codes, so for psuedo-exhaustive generation we must first exhaustively generate all cyclic codes in $\mathbb{F}_2$. True exhaustive generation would be extremely computationally expensive and impossibly time consuming, so to shorten the workload, we used the cyclic partition algorithm described in \cite{GenASR} to instead generate a representative code from each %equivalence class as defined by the rules for code equivalence.

The generalized ASR algorithm is based on the notion of equivalent codes. 

\begin{definition} Two linear codes are  equivalent if one can be obtained from the other by any combinations of the following transformations:

\begin{enumerate}
    \item A permutation of coordinates.
    \item Multiplication of elements in a fixed position by a non-zero scalar in $\mathbb{F}_q$.
    \item Applying an automorphism of $\mathbb{F}_q$ to each component of the vectors.
\end{enumerate}
\end{definition}

If only the first transformation is used then the resulting codes are called permutation equivalent. This is a very important special case and in fact, for binary codes it is the only type of code equivalence that is possible. 

%\begin{definition} If one code can be obtained from the other using only a permutation of coordinates, then the codes are permutation equivalent.
%\end{definition}

The algorithm given in \cite{cycliceq} allows us to partition cyclic codes of a given length into equivalence classes, and then choose one code from each class, significantly reducing the computational workload and allowing us to quickly move on to QC construction. 

The algorithm for a cyclic code over $\mathbb{F}_2$ is as follows:
\begin{enumerate}
    \item Start with a length $n$.

    \item Write $n$ in the form $n'2^t$ such that $n'$ is not divisible by $2$.

    \item Generate cyclotomic cosets of $q$ mod $n'$.

   % \item From the cyclotomic cosets, pick cosets whose union forms the set $\{0,1,2,\dots,{n-1}\}$. \hl{This step is not clear to me.}

    \item Generate all multisets of the cosets such that each coset can be repeated up to $2^t$ times. 

    \item Check for linear maps between the generated multisets of the same size. If a linear map exists between two multisets, then we know that the codes defined by them are equivalent so we eliminate one of them.

    \item Find $\beta$, a primitive $n'$-th root of unity over $\mathbb{F}_2$.
    
    \item Use the remaining multisets and $\beta$ to obtain generator polynomials such that a multiset $w = \{w_0,w_1,...,w_i\}$ generates the polynomial $g_w(x) = (x - \beta^{w_0})(x - \beta^{w_1})\cdots (x - \beta^{w_i})$.
\end{enumerate}

%\hl{Good news! Actually for binary codes, it will catch everything!}
As explained in \cite{cycliceq}, while in general it is possible for this algorithm to fail to distinguish between some equivalent codes, over the binary field it is guaranteed to completely partition the code space. Checking equivalence of codes once they have been generated is known to be equivalent to the graph isomorphism problem, which is believed to be NP-Intermediate \cite{EquivComp}. Magma software has a function for this task for general linear codes but it does not always work and in many cases it takes too long to finish. Our algorithm that is specifically for cyclic codes is much more efficient. We refer the reader to \cite{cycliceq} for more on the details and performance of our algorithm. Having this algorithm to quickly produce cyclic codes, we could then move on to using the ASR search algorithm to generate QC codes.

%\medskip 

\subsection{The ASR algorithm. \\}

Given a generator polynomial $p(x) = p_0 + p_1x + \cdots + p_{m-1}x^{m-1}$ of a cyclic code $C$ of length $m$ over finite field $\mathbb{F}_q$, $C$ has a generator matrix of the following form:

\[
G=
  \begin{bmatrix}
    p_0 & p_{1} & p_{2} & \cdots & p_{m-1} \\
    p_{m-1} & p_{0} & p_{1} & \cdots & p_{m-2} \\
    p_{m-2} & p_{m-1} & p_{0} & \cdots & p_{m-3} \\
    \vdots & \vdots &  \vdots &  &  \vdots  \\
    p_{m-k+1} & p_{m-k+2} & p_{m-k+2} & \cdots & p_{m-k}
  \end{bmatrix}
\]

Such a matrix is called a circulant matrix \cite{Twistulant}. As a generalization of cyclic codes, a generator matrix of a QC code consists of blocks of circulant matrices. In general, a generator matrix of an $\ell$-QC code has the following form
\[
G=
  \begin{bmatrix}
    G_{11} & G_{12} & \cdots & G_{1\ell} \\
    G_{21} & G_{22} & \cdots  & G_{2\ell}\\
    \vdots & \vdots &   & \vdots  \\
    G_{r1} & G_{r2} & \cdots & G_{r\ell}
    
  \end{bmatrix}
\]
where each $G_{ij}$ is a circulant matrix corresponding to a cyclic code. Such a code is called an $r$-generator QC code. \cite{Twistulant} The case  $[G_1 \; G_2 \; ... \; G_{\ell}]$ gives 1-generator QC codes, which is the case we have considered in this work. 

\medskip 

We begin the ASR search algorithm by taking one generator $g(x)$ of a cyclic code of length $m$ from each equivalence class. Then, we construct the generator of an $\ell$-QC code in the form

$$( f_1(x)g(x), f_2(x)g(x), . . . , f_l(x)g(x)),$$

\smallskip 

\noindent where all $f_i(x)$ are chosen arbitrarily from $\mathbb{F}_q[x]/ \langle x^m-1\rangle$ such that they are relatively prime to $h(x)$, the check polynomial of the cyclic code generated by $g(x)$, and $\deg(f_i(x))<\deg(h(x))$. The following theorem is the basis of the ASR algorithm.

\begin{theorem}
\emph{\cite{ASR}} Let $C$ be a $1$-generator $\ell$-QC code over $\Fq$ of length $n = m\ell$ with a generator $G(x)$ of the form:

$$
G(x) = \left( f_1(x)g(x), f_2(x)g(x), . . . , f_l(x)g(x)\right),
$$

\smallskip

\noindent where $x^m - 1  = g(x)h(x)$ and for all $i=1 ..., \ell$, $gcd(h(x),f_i(x))=1$ . Then, $C$ is an $[n,k,d']_q$-code where $k=m-deg(g(x)),$ and $d' \geq \ell \cdot d$, $d$ being the the minimum distance of the cyclic code $C_g$ of length $m$ generated by $g(x)$.

\end{theorem}

\section{New Binary QC Codes}

Using this search method, we have been able to generate 113  binary QC codes with the following features.

\begin{enumerate}
 
     \item Every one of these codes is new among the  class of binary QC codes according to the database \cite{qcdatabase}.
     
    \item Each of these codes has the same parameters as BKLCs in \cite{database}.
   
    \item In many cases, the BKLCs in the database (\cite{database}) have indirect, multi-step constructions so it is more efficient and desirable to obtain them in the form of QC codes instead. All of the codes we present are QC and in many cases the corresponding codes in \cite{database} do not have simple constructions.

    \item A number of our codes have additional desirable properties such as  being self-orthogonal, reversible, and linear complementary dual (LCD).
\end{enumerate}

 The following table lists the parameters, properties, and generators of these new QC codes. The generators are listed by their coefficients in base 8 for a compact representation. For example, consider the length $n=70$ code in Table 2 below whose generator is $1 + x^2 + x^3 + x^4$. Its coefficients are $10111$ in increasing powers of $x$ from left to right. We break this up into blocks of three, so $101$,$110$. These blocks are then converted to base 8, reading left to right, so they become $53$. The number of $f$ polynomials corresponds to the number of blocks, or index ($\ell$) of the QC code, and $m$ by $\ell$ gives $n$, the total length. For this particular example, $\ell=2$, hence the block length is $m=35$. This means $g(x)$ is a divisor of $x^{35}-1$. 

%\newpage

\newpage 

\makeatletter
\def\old@comma{,}
\catcode`\,=13
\def,{%
  \ifmmode%
    \old@comma\discretionary{}{}{}%
  \else%
    \old@comma%
  \fi%
}
\makeatother

\begin{center}

\begin{longtable}{l|l|p{0.70\linewidth}}
\caption{New Binary QC Codes that are LCD}\\
$[n,k,d]_q$ & $g$ & $[f_1,...,f_\ell]$\\
\hline&&\\
$ [195,38,58]_2$ & $3$ & $[ 6135332413622, 5511163061413, 0760162305021, 
41167113347, 1304422614561 ]$\\
$ [172,42,46]_2$ & $3$ & $[ 16305255301422, 12552443046071, 26367761131227, 
56712075265061 ]$\\
$ [165,32,50]_2$ & $3$ & $[ 7451352027, 7530304175, 16125740433, 70670753002,42255573462 ]$\\
$ [164,40,44]_2$ & $3$ & $[ 0711720165373, 4162676126303, 731245452506, 
4301603434632 ]$\\
$ [156,36,44]_2$ & $11$ & $[ 377230104735, 441101605347, 135525223756, 
034664137676 ]$\\
$ [153,43,38]_2$ & $727$ & $[ 03362056173102, 273542165074071, 507007322260431 ]$\\
$ [141,46,32]_2$ & $3$ & $[ 340604477037257, 5555546265363011, 1447675154362301 ]$\\
$ [124,30,36]_2$ & $3$ & $[ 3023466614, 104135571, 6774200266, 4032715624 ]$\\
$ [122,60,20]_2$ & $3$ & $[ 22027750476404545177, 45756235144775354244 ]$\\
$ [120,32,32]_2$ & $104$ & $[ 3211420427, 0605045201, 45670476662 ]$\\
$ [116,28,34]_2$ & $3$ & $[ 442732531, 253642371, 2645552751, 4024662711 ]$\\
$ [114,54,20]_2$ & $11$ & $[ 350273450337664702, 105500762001021221 ]$\\
$ [111,36,26]_2$ & $3$ & $[ 546217552016, 4415543004, 640326514123 ]$\\
$ [110,50,20]_2$ & $14$ & $[ 4002671571611246, 0312571544502463 ]$\\
$ [110,40,2]_2$ & $730471$ & $[ 42213112722401, 0752260652054 ]$\\
$ [108,48,20]_2$ & $101$ & $[ 3300216514056443, 7000312523564625 ]$\\
$ [108,32,28]_2$ & $12$ & $[ 74160521111, 45604632562, 40250112373 ]$\\
$ [105,34,26]_2$ & $3$ & $[ 742252523401, 314437023031, 070422111261 ]$\\
$ [105,29,28]_2$ & $771$ & $[ 6162561672, 6651775572, 1064436731 ]$\\
$ [104,24,32]_2$ & $5$ & $[ 57465517, 0174333, 31361042, 64356021 ]$\\
$ [100,40,20]_2$ & $1002$ & $[ 3367605450137, 2264022063455 ]$\\
$ [99,32,24]_2$ & $3$ & $[ 65725410163, 17752117321, 5251344657 ]$\\
$ [99,21,32]_2$ & $57731$ & $[ 2433502, 5112553, 6660032 ]$\\
$ [94,46,16]_2$ & $3$ & $[ 17570216336424, 6073617230441121 ]$\\
$ [93,30,24]_2$ & $3$ & $[ 7475563176, 4415177161, 234165126 ]$\\
$ [88,20,28]_2$ & $5$ & $[ 4642143, 1264541, 7704431, 247022 ]$\\
$ [84,24,24]_2$ & $12$ & $[ 55575737, 02203365, 31403363 ]$\\
$ [78,24,22]_2$ & $5$ & $[ 27512541, 02121473, 60544261 ]$\\
$ [58,28,12]_2$ & $3$ & $[ 4127557501, 402073244 ]$\\
$ [52,24,12]_2$ & $5$ & $[ 7360021, 5267555 ]$
\label{tab:my_label}
\end{longtable}
    
\end{center}

\begin{table}
    \centering
    \caption{New Binary QC Codes that are Self-Orthogonal}
    \begin{tabular}{l|l|p{0.70\linewidth}}
$[n,k,d]_q$ & $g$ & $[f_1,...,f_\ell]$\\
\hline&&\\
$ [213,35,68]_2$ & $7120103605521$ & $[ 111124074763, 432070306671, 774512423733
]$\\
$ [158,39,44]_2$ & $35441216370232$ & $[ 5172674573162, 1174545113363 ]$\\
$ [93,15,36]_2$ & $525412$ & $[ 73036, 34767, 46131 ]$\\
$ [70,31,16]_2$ & $53$ & $[ 0215201037, 17453360511 ]$\\
$ [69,22,20]_2$ & $3$ & $[ 12325661, 6003045, 10405 ]$\\
$ [66,20,20]_2$ & $5$ & $[ 3343631, 027677, 0516553 ]$\\
    \end{tabular}
    \label{tab:my_label}
\end{table}
\begin{table}{
    \centering
    \caption{New Binary QC Codes that are Self-Orthogonal and Reversible}
    
    \begin{tabular}{l|l|p{0.70\linewidth}}
$[n,k,d]_q$ & $g$ & $[f_1,...,f_\ell]$\\
\hline&&\\
$ [70,30,16]_2$ & $14$ & $[ 1367566016, 5714137543 ]$\\
$ [52,25,12]_2$ & $3$ & $[ 156307741, 752562251 ]$\\
    \end{tabular}
    \label{tab:my_label}
}
\end{table}

\newpage

In addition to the additional properties that they posses, these new QC codes are usually better than the  BKLCs currently listed in the database \cite{database} for the reason that their constructions are far simpler. A QC code is more desirable than an arbitrary linear code for many reasons. It has a well understood  algebraic structure and  its generator matrix is determined by its first row alone. This property is being exploited in some cryptosystems that are based in coding theory to reduce the key sizes in McEliece type crytosystems (\cite{{smallkeys}}).  In comparison, the BKLCs in \cite{database} with the same parameters can have far more steps to construct the code. For example, we found a $[213,35,68]_2$ code that has the same parameters as the comparable BKLC, as well as being self orthogonal and reversible. Being a QC code, it has a single step construction. The current record holder for the same length, dimension and field in \cite{database} on the other hand has a 17-step  construction to achieve the same parameters and lacks any additional properties. In many case, the codes presented are far simpler and desirable with the same parameters as BKLCs.  

The codes that have additional properties are already  listed in the tables above. Any new codes that are not  listed in the tables above have their parameters recorded below. For the sake of space, we do not write down their generators. They are available from the authors. Moreover, these codes have been added to the database \cite{qcdatabase} and  their generators are available there as well.

\begin{multicols}{4}
{\renewcommand\labelitemi{}
\begin{itemize}[leftmargin=*]
\item $ [218,37,68]_2$
\item $ [200,37,60]_2$
\item $ [190,37,56]_2$
\item $ [175,34,52]_2$
\item $ [168,41,44]_2$
\item $ [160,38,44]_2$
\item $ [160,37,44]_2$
\item $ [160,30,50]_2$
\item $ [153,42,38]_2$
\item $ [150,45,36]_2$
\item $ [150,44,36]_2$
\item $ [144,45,34]_2$
\item $ [138,45,32]_2$
\item $ [132,41,32]_2$
\item $ [132,34,36]_2$
\item $ [132,31,38]_2$
\item $ [130,53,24]_2$
\item $ [128,52,24]_2$
\item $ [128,51,24]_2$
\item $ [128,31,36]_2$
\item $ [124,51,24]_2$
\item $ [124,50,24]_2$
\item $ [124,37,32]_2$
\item $ [120,58,20]_2$
\item $ [120,49,24]_2$
\item $ [120,47,24]_2$
\item $ [120,46,24]_2$
\item $ [120,39,28]_2$
\item $ [120,33,32]_2$
\item $ [112,52,20]_2$
\item $ [112,51,20]_2$
\item $ [112,50,20]_2$
\item $ [112,47,22]_2$
\item $ [112,27,32]_2$
\item $ [112,23,36]_2$
\item $ [110,45,22]_2$
\item $ [110,41,24]_2$
\item $ [108,49,20]_2$
\item $ [108,47,20]_2$
\item $ [108,46,20]_2$
\item $ [108,35,26]_2$
\item $ [108,31,28]_2$
\item $ [108,26,32]_2$
\item $ [105,31,28]_2$
\item $ [102,42,20]_2$
\item $ [102,32,26]_2$
\item $ [100,42,20]_2$
\item $ [100,41,20]_2$
\item $ [96,39,20]_2$
\item $ [96,38,20]_2$
\item $ [96,37,20]_2$
\item $ [96,36,20]_2$
\item $ [96,31,24]_2$
\item $ [96,30,24]_2$
\item $ [92,35,20]_2$
\item $ [92,34,20]_2$
\item $ [90,40,18]_2$
\item $ [90,34,20]_2$
\item $ [90,28,24]_2$
\item $ [90,27,24]_2$
\item $ [88,32,20]_2$
\item $ [84,31,20]_2$
\item $ [84,30,20]_2$
\item $ [84,23,24]_2$
\item $ [81,26,22]_2$
\item $ [80,37,16]_2$
\item $ [80,28,20]_2$
\item $ [72,31,16]_2$
\item $ [70,29,16]_2$
\item $ [64,29,14]_2$
\item $ [64,25,16]_2$
\item $ [64,24,16]_2$
\item $ [60,26,14]_2$
\item $ [56,19,16]_2$
\item $ [48,21,12]_2$

\end{itemize}
}
\end{multicols}

\section{A ConstructionX Method For New Codes From QC Codes}

In this section we examine the construction of new good codes from existing QC codes using ConstructionX. This method was inspired by similar work in \cite{constx}, however we have generalized it and expanded on the details of the methodology.

ConstructionX is a method of creating new codes from existing good codes (\cite{constX},\cite{bible}). Given a code $C_1$ with parameters $[n_1,k_1,d_1]$, a subcode $C_2$ of it  with parameters $[n_1,k_1 - b, d_2]$, and a third code $C_3$ with parameters $[n_2,b,d_3]$, ConstructionX divides $C_2$ into a union of cosets of $C_1$ and attaches a different codeword of $C_3$ to each coset. This results in a new code, $C$, which has parameters $[n,k,d]$ such that $n=n_1+n_3$, $k=k_1$, and $d_2 \geq d \geq \min\{d_2,d_1+d_3\}$. Furthermore, $C$ has a generator matrix of the form 
$$
\begin{bmatrix}
G_1^* & G_3\\
G_2 & 0
\end{bmatrix},
$$
where $G_2$ is a generator matrix of $C_2$, $G_3$ is the generator matrix of $C_3$ and the rows of $G_1^*$ are a set of linearly independent vectors of $C_1$ that are not in $C_2$ such that
$$
\begin{bmatrix}
G_1^*\\
G_2
\end{bmatrix}
$$ generates $C_1$.

The main problem in applying ConstructionX is finding the best possible $C_2$ while keeping $b$ small. Since $d_2$ is an upper bound, we want to maximize the minimum distance of $C_2$, however, we also want a good minimum distance of $C_3$ in order to maximize the lower bound, which ConstructionX does not often exceed. In fact, none of the ConstructionX codes presented in this paper exceed the lower bound, and each of them either have $d = d_1+d_3$ or $d = d_2 = d_1+d_3$. None have $d = d_2 \not \eq d_1 + d_3$ or $d_2 \not \eq d \not \eq d_1+d_3$. Thus, want to pick a $C_3$ of lower dimension, which will generally increase $d_3$. The challenge then becomes finding a good $C_2$ of higher dimension, or in other words setting $b$ to a small value. In our searches we examined specifically $1 \leq b \leq 6$, and did not find any new BKLCs with $b \geq 3$.

In \cite{constx}, new record breaking codes over $\mathbb{F}_3$ were found by looking at QC subcodes of good QC codes. We implemented an algorithm to examine all QC subcodes of a given dimension for a good QC code and choose the best one for use in ConstructionX, and additionally did the same for supercodes.

%The following results are well known in coding theory:

Our method is based on the following observations:

\begin{itemize}
    \item Let $C$ be a 1-generator QC  code  generated by $(gf_1,gf_2,...,gf_{\ell})$. Then for each divisor $p$ of $g$, the 1-generator QC code generated by $(g'f_1,g'f_2,...,g'f_{\ell})$ is a supercode of $C$, where $g'=\frac{g}{p}$
    
    \item Let  $C$ be a 1-generator QC  code  generated by $(gf_1,gf_2,...,gf_{\ell})$ where $x^m-1=gh$. Then for each divisor $p$ of $h$, the 1-generator QC code generated by $(pgf_1,pgf_2,...,pgf_{\ell})$ is a subcode of $C$.
\end{itemize}

\begin{comment}
\begin{theorem}
Let C be a Quasi-Cyclic $[n,k,d]_q$ code generated by the set $fs = \{gf_1,gf_2,...,gf_\ell\}$ of polynomials over $\mathbb{F}_q$. Then there is a relationship between QC supercodes of $C$ and factors of $g$ such that for each $p$ that is a factor of $g$, there exists a QC supercode with a set of generator polynomials of the form $$fs_{p} = \{\frac{g}{p}f_1,\frac{g}{p}f_2,...,\frac{g}{p}f_\ell\}.$$
\end{theorem}

\begin{theorem}
Let C be a Quasi-Cyclic $[n,k,d]_q$ code generated by the set $fs = \{gf_1,gf_2,...,gf_\ell\}$ of polynomials over $\mathbb{F}_q$. Then the check polynomial of $C$ is $${h = \frac{x^m-1}{g}},$$ and there is a relationship between QC supercodes of $C$ and factors of $h$ such that for each $p$ that is a factor of $h$, there exists a QC subcode with a set of generator polynomials of the form $$fs_p = \{pgf_1,pgf_2,...,pgf_\ell\}.$$
\end{theorem}
\end{comment}

Using these facts, we are able to quickly examine QC sub and supercodes for a given QC code. In our searches, we used QC codes with the parameters of BKLCs as our initial QC code, as well as the binary QC codes outlined above. Codes were pulled from the database at \cite{database} and checked for being Q. If they were in fact QC codes, we tested them using this algorithm. While $g$ may be found through examining the inputs of the ASR algorithm, it is more generally found by taking the greatest common divisor of the set of generator polynomials. Our algorithm for supercode examination in psuedocode is as follows:

\bigskip

\begin{algorithm}[H]
\SetAlgoLined
 Input: $fs = [f_1,\dots,f_{\ell}]$\;
 Input: $b$\;
 $g = \gcd(fs)$\; 
 C = QCcode(fs)\;
 \While{Factors of $g$ remain}{
  factor = The next factor of $g$\;
  \If{Degree(factor) $\not \eq $ b}{
   continue\;
   }
   newfs = [$\frac{f}{factor}$ for f in fs]\;
   superC = QCcode(newfs)\;
  \If{MinimumDistance(Best) $<$ MinimumDistance(superC)}{
   Best = SuperC\;
  }
  \For{length to max}{
    C3 = BKLC(length,b)\;
    CX = ConstructionX(C,Best,C3)\;
    print(CX)\;
  }
 }
 \KwResult{New high minimum distance codes from ConstructionX}
 \caption{Finding ConstructionX Codes from Good QC Codes}
\end{algorithm}

\section{New BKLCs From ConstructionX}

Using this method, we found 35 new record breaking linear codes over $\mathbb{F}_2$, $\mathbb{F}_3$, $\mathbb{F}_4$ and $\mathbb{F}_5$ according to the database at \cite{database}, including one new QC code which is a BKLC that was found incidentally.

\begin{theorem}
There exist linear codes with the following parameters: $[98,30,26]_2$,  $[97,30,25]_2$, $[99,31,26]_2$,  $[98,31,25]_2$, $[176,51,41]_2$, $[177, 52, 41]_2$, $[177,51,42]_2$, $[178, 52, 42]_2$, $[112,23,46]_3$, $[100, 28, 35]_3$, $[101,26,37]_3$, $[105, 31, 35]_3$,  $[106,23,43]_3$, $[107, 23, 43]_3$, $[108, 31, 36]_3$, $[113,23,47]_3$,  $[114,23,48]_3$,  $[115,23,48]_3$, $[141,26,59]_3$, $[164,27,70]_3$, $[166,27,71]_3$, $[167,27,72]_3$, $[168,27,72]_3$, $[169,26,74]_3$, $[170,26,75]_3$, $[170,27,73]_3$, $[171,26,75]_3$, $[172,26,75]_3$, $[190, 24, 88]_3$, $[217,14,121]_3$, $[218,14,122]_3$, $[219,14,123]_3$, $[140,18,75]_4$, $[143,19,75]_4$, $[81, 18, 40]_5$
\end{theorem}

The parameters of the codes and the corresponding supercodes used in the construction of the new codes are:

\begin{table}[htb]
    \centering
    \caption{Record Breaking ConstructionX Codes}
    \begin{tabular}{l|l|l|l}
New Code & Original Code & Supercode & Third Code\\

\hline&&\\
$[98,30,26]_2$ & $[96,29,26]_2$ & $[96,30,24]_2$ & $[2,1,2]_2$\\
$[97,30,25]_2$ & $[96,29,26]_2$ & $[96,30,24]_2$ & $[1,1,1]_2$\\
$[99,31,26]_2$ & $[96,29,26]_2$ & $[96,31,24]_2$ & $[3,2,2]_2$\\
$[98,31,25]_2$ & $[96,29,26]_2$ & $[96,31,24]_2$ & $[2,2,1]_2$\\
$[177, 52, 41]_2$ & $[170,48,42]_2$ & $[170,52,38]_2$ &  $[7,4,3]_2$\\
$[178, 52, 42]_2$ & $[170,48,42]_2$ & $[170,52,38]_2$ &  $[8,4,4]_2$\\
$[100, 28, 35]_3$ & $[99,27,35]_3$ & $[99,28,34]_3$ & $[1,1,1]_3$\\
$[101,26,37]_3$ & $[99,25,37]_3$ & $[99,26,35]_3$ & $[2,1,2]_3$\\
$[105, 31, 35]_3$ & $[104,30,35]_3$ & $[104,31,34]_3$ & $[1,1,1]_3$\\
$[107, 23, 43]_3$ & $[104,21,43]_3$ & $[104,23,41]_3$ & $[3,2,2]_3$\\
$[108, 31, 36]_3$ & $[104,28,37]_3$ & $[104,31,34]_3$ & $[4,3,2]_3$\\
$[113,23,47]_3$ & $[112,22,48]_3$ & $[112,23,46]_3$ & $[1,1,1]_3$\\
$[114,23,48]_3$ & $[112,22,48]_3$ & $[112,23,46]_3$ & $[2,1,2]_3$\\
$[115,23,48]_3$ & $[112,22,48]_3$ & $[112,23,46]_3$ & $[3,1,3]_3$\\
$[164,27,70]_3$ & $[160,24,72]_3$ & $[160,27,68]_3$ & $[4,3,2]_3$\\
$[166,27,71]_3$ & $[160,24,72]_3$ & $[160,27,68]_3$ & $[6,3,3]_3$\\
$[167,27,72]_3$ & $[160,24,72]_3$ & $[160,27,68]_3$ & $[7,3,4]_3$\\
$[168,27,72]_3$ & $[160,24,72]_3$ & $[160,27,68]_3$ & $[8,3,5]_3$\\
$[169,26,74]_3$ & $[160,22,75]_3$ & $[160,26,69]_3$ & $[9,4,5]_3$\\
$[170,26,75]_3$ & $[160,22,75]_3$ & $[160,26,69]_3$ & $[10,4,6]_3$\\
$[170,27,73]_3$ & $[160,23,73]_3$ & $[160,27,67]_3$ & $[10,4,6]_3$\\
$[171,26,75]_3$ & $[160,22,75]_3$ & $[160,26,69]_3$ & $[11,4,6]_3$\\
$[172,26,75]_3$ & $[160,22,75]_3$ & $[160,26,69]_3$ & $[12,4,6]_3$\\
$[190, 24, 88]_3$ & $[182,21,88]_3$ & $[182,24,83]_3$ & $[8,5,3]_3$\\
$[143, 19, 75]_4$ & $[140,18,75]_4$ & $[140, 19, 72]_4$ & $[3,1,3]_4$\\
$[81, 18, 40]_5$ & $[78,16,40]_5$ & $[78,18,37]_5$ & $[3,2,2]_5$
    \end{tabular}
    \label{tab:my_label}
\end{table}

 We also found a few ternary record breakers by examining subcodes as opposed to supercodes.

\begin{table}[htb]
    \centering
    \caption{Record Breaking ConstructionX Codes}
    \begin{tabular}{l|l|l|l}
New Code & Original Code & Subcode & Third Code\\

\hline&&\\
$[141,26,59]_3$ & $[140, 26, 58]_3$ & $[140, 25, 59]_3$ & $[1,1,1]_3$\\
$[217,14,121]_3$ & $[208,14,117]_3$ & $[208,9,126]_3$ & $[9,5,4]_3$\\
$[218,14,122]_3$ & $[208,14,117]_3$ & $[208,9,126]_3$ & $[10,5,5]_3$\\
$[219,14,123]_3$ & $[208,14,117]_3$ & $[208,9,126]_3$ & $[11,5,6]_3$\\
    \end{tabular}
    \label{tab:my_label}
\end{table}

Finally, more record breakers were found by modifying the codes in Tables 4 and 5.

\newpage

\begin{table}[htb]
    \centering
    \caption{New Record Breakers By Modification}
    \begin{tabular}{l|l|l}
New Code & Original Code & Modification Method\\

\hline&&\\
$[177,51,42]_2$ & $[177,52,41]_2$ & Expurgation\\
$[176,51,41]_2$ & $[177,52,41]_2$ & Shorten at position 169\\
$[106,23,43]_3$ & $[107,23,43]_3$ & Puncture at position 106\\
$[140,18,75]_4$ & $[143, 19, 75]_4$ & Shorten at positions 141, 142 and 143\\
    \end{tabular}
    \label{tab:my_label}
\end{table}

The table below presents the generators for the original and super/subcodes that were used as components of the new ConstructionX code. Each of the codes in the table corresponds to a set of parameters found in Table 4 or Table 5.

The $\mathbb{F}_2$ codes are presented in the same way as the previous tables, and the $\mathbb{F}_3$ codes are presented with the same algorithm, but instead of converting 3 places to base 8, it converts 2 places to base 9. The $\mathbb{F}_4$ codes are presented unabridged and with $b=a^2$. $\mathbb{F}_5$ codes are also presented unabridged.

Since there are a few codes in this table which share parameters while  being non-equivalent, each code is presented adjacent to its respective sub/supercode, and in approximately the same order as the table in which they appear. One of the QC supercodes, with parameters $[112,23,46]_3$, in the table below is also a record breaker in its own right.

\begin{table}[htb]
    \centering
    \caption{QC Codes Which Generate Record Breaking ConstX Codes}
    \begin{tabular}{l|l}
Parameters & Generators\\
\hline&\\

$[96,29,26]_2$ & $[71, 64113173343, 27771046431]$\\
$[96,30,24]_2$ & $[5, 24076056631, 6252773246]$\\
$[96,31,24]_2$ & $[3,6305574556,2641521632]$\\
$[170,48,42]_2$ & $[1000000000000000377323447615,4135600657027030415272654623]$\\
$[170,52,38]_2$ & $[341603416034160314066672363,427200026166224122662430431]$\\
$[140, 26, 58]_3$ & $[75656265543767850565651,6273308468022430087634822263154874]$\\
$[140, 25, 59]_3$ & $[83434633507657031443433,36887111402772710175802677383402711]$\\
$[99,27,35]_3$ & $[03000000000003141,5151832674585001,4201041431785337]$\\
$[99,28,34]_3$ & $[0688888888888505,186426033663544,5884464650276512]$\\
$[99,25,37]_3$ & $[000000100000385,51172578486836861,65173507073567432]$\\
$[99,26,35]_3$ & $[000000888888521,1843867363731323,3502668787557716]$\\
$[104,30,35]_3$ & $[00000000001000034635424162,63878246257051483836350245]$\\
$[104,31,34]_3$ & $[0000000000888885626614647,31773813867864631757554811]$\\
$[112,22,48]_3$ & $[225217636715327031,3855753628142726518311446808]$\\
$[112,23,46]_3$ & $[48608726228658785,6355351383058703507508157342]$\\
$[104,21,43]_3$ & $[00000000036740414327721684,13231000037464840407461361]$\\
$[104,23,41]_3$ & $[0000000006051168412467604,2863555552403232277082171]$\\
$[104,28,37]_3$ & $[30000000000000774187451744,35747612345623613624005564]$\\
$[104,31,34]_3$ & $[654720654720653833336071,634261474208860240816081]$\\
$[160,24,72]_3$ & $[1000000000007321057462270530227344544615,$\\
&$6588363577683773486427864331370312740533]$\\
$[160,27,68]_3$ & $[856085608560785553784776448685835748281,$\\
&$350237067675375315443715352658617323401]$\\
$[160,22,75]_3$ & $[000300000004061832741081527113005224724,$\\
&$763324150001553440888665474088827327074]$\\
$[160,26,69]_3$ & $[0003628805165221013063606507063164863,$\\
&$7841540000011055215802127766367524533]$\\
$[160,23,73]_3$ & $[0000003000068511234582485684775686524412,$\\
&$6326771628047648801417446130222227241818]$\\
$[160,27,67]_3$ & $[0000007032060245487053016568864183237,$\\
&$54702682323740810856185340274102477412]$\\
$[208,14,117]_3$ & $[0748724866871624680580367184142241783262013113142526,$\\
&$1245403380035706885414525207425072611683628176418272]$\\
$[208,9,126]_3$ &  $[0431024638315531438181714874147403865036051473417760472,$\\
&$707452166345144686117316063134645810071144055574570547]$\\
$[140,18,75]_4$ & $[101,aababba1aa1b11b0baa,111110abab1a100b1a0b,b1b10b010b01ba1100ab,$\\
& $1110a0010aaa01bab10a,b1bab0baaaa0a1bbaabb,ba0baabbb0a0a1babb10]$\\
$[140, 19, 72]_4$ & $[11,a0b1a1ba0ab010bb0a,101011b0a10abbb01bb,ba100bbaa110b10111b,$\\
& $1011bbbaa0a001a0baa,ba1b00b1b1bb10b0a0b,b11a0a1a11bb10b1a10]$\\
$[78,18,40]_5$ & $[12312024143330311210411103220134021044,$\\
&$111341132034241330331232130030204321433]$\\
$[78,18,37]_5$ & $[111424141431404200144344223204410104,$\\
&$1003123311124341314340434104421210103]$\\
    \end{tabular}
    \label{tab:my_label}
\end{table}


\clearpage

\newpage \begin{thebibliography}{}


\bibitem{magma} Magma computer algebra system, online, \url{http://magma.maths.usyd.edu.au/}

\bibitem{database} Grassl M.: Code Tables: Bounds on the parameters of of codes, online, \url{http://www.codetables.de/}

\bibitem{qcdatabase} Chen, E.: Quasi-Cyclic Codes: Bounds on the parameters of of QC codes, online, \url{http://www.tec.hkr.se/~chen/research/codes/qc.htm}




\bibitem{cycliceq}  Aydin, N.,   VandenBerg R. O., A New Algorithm for Equivalence of Cyclic Codes and Its Applications,  preprint, arXiv:2107.00159, 2021 - arxiv.org

\bibitem{EquivComp} E. Petrank and R. M. Roth, "Is code equivalence easy to decide?," in IEEE Transactions on Information Theory, vol. 43, no. 5, pp. 1602-1604, Sept. 1997, doi: 10.1109/18.623157.


\bibitem{GenASR}Aydin N., Lambrinos, J., VandenBerg, R. O.: On Equivalence of Cyclic Codes, Generalization of a Quasi-Twisted Search Algorithm, and New Linear Codes.  Design Code Cryptogr. 87, 2199-2212 (2019)

\bibitem{constx} Daskalov, Rumen \& Hristov, Plamen. (2017). Some new ternary linear codes. JOURNAL OF ALGEBRA COMBINATORICS DISCRETE STRUCTURES AND APPLICATIONS. 4. 227-234.

\bibitem{ackerman2011} R. Ackerman, and N. Aydin, New quinary linear codes from quasi-twisted codes and their duals, Applied Mathematics Letters, Vol 24, No 4, pp 512-515, April 2011.

\bibitem{daskalov2003} R. Daskalov and P. Hristov, New quasi-twisted degenerate ternary linear codes, IEEE Transactions on Information Theory, vol. 49, no. 9, pp. 2259-2263, Sept. 2003, doi: 10.1109/TIT.2003.815798.

\bibitem{ASR}Aydin, N., Siap, I., Ray-Chaudhuri, D.: The structure of 1-generator quasi-twisted codes and new linear codes. Design Code Cryptogr.  24, 313-326 (2001)

\bibitem{Twistulant} Some generalizations of the ASR search algorithm for quasitwisted codes Nuh Aydin, Thomas H. Guidotti, Peihan Liu, Armiya S. Shaikh and Robert O. VandenBerg (2020)

%\bibitem{mt}Aydin N., Halilovic, A.: A generalization of quasi-twisted codes: Multi-twisted codes. Finite Fields Appl.  45, 96-106 (2017)

\bibitem{gf7}Aydin N., Connolly, N., Grassl, G.: Some results on the structure of constacyclic codes and new linear codes over GF(7) from quasi-twisted codes. Adv. Math. Commun. 11, 245-258 (2017)



\bibitem{NPhard}Vardy, A.: The intractability of computing the minimum distance of a code. IEEE Trans. Inform. Theory. 43 1757-1766 (1997)


 
%\bibitem{Quantumoriginal1}Steane, A. M.: Error correcting codes in quantum theory. Phys. Rev. Lett. 77, 793-797 (1996)

%\bibitem{Quantumoriginal2}Calderbank, A. R., Shor, P. W.: Good quantum error-correcting codes exist. Phys. Rev. A. 54, 1098-1106 (1996)

\bibitem{Quantumoriginal3}Calderbank, A. R., Rains, E. M., Shor, P. W., Sloane, N. J. A.: Quantum error correction via codes over GF(4). IEEE Trans. Inform. Theory. 44, 1369-1387 (1998)

\bibitem{mtgood} Aydin, N., Guidotti, T., Liu, P.: Good Classical and Quantum Codes from Multi-Twisted Codes, arxiv.org, 2008.07037, (2020)

%\bibitem{oto}Parra-Avila, B., Permouth, S., Szabo, S.: Dual Generalizations of the Concept of Cyclicity of Codes. Adv. Math. Commun. 3, 227-234 (2009)

%\bibitem{pc2011}Matsuoka, M.: $\theta$-Polycyclic codes and $\theta$-sequential codes over finite field. Int. J.  Algebra. 5, 65-70 (2011)

%\bibitem{pc2016}Alahmadi, A., Dougherty S., Leroy, A., Solé, P.: On the duality and the direction of polycyclic codes. Adv. Math. Commun. 10, 921-929 (2016)

%\bibitem{pc2020}Fotue-Tabue, A., Martinez-Moro, E., Blackford J. T.: On polycyclic codes over a finite chain ring, Adv. Math. Commun. 14, 455-466 (2020)

\bibitem{mas}Massey, J.: Reversible Codes. Information and Control. 7, 369-380 (1964)

\bibitem{LCD} J.L. Massey, Linear codes with complementary duals, Discrete Math. 106/107 (1992) 337-342

\bibitem{LCDuse} Lu, L.,Li, R.,Fu, Q.,Xuan, Chen.,Ma, Wenping.: Optimal Ternary Linear Complementary Dual Codes. arXiv:2012.12093v2 [cs.IT] (2020)
 
%\bibitem{Lid}Rudolf, L., Harald, N.: Introduction to finite fields and their applications. Cambridge University Press (1986)

%\bibitem{dna} Abualrub, T., Ghrayebb, A., Zeng, X.: Construction of cyclic codes over GF(4) for DNA computing. J. Franklin Inst. 343, 448-457 (2006)

%\bibitem{dna2}Oztas E. S., Yildiz, B., Siap, I.: A novel approach for constructing reversible codes and applications to DNA codes over the ring $F_{2}[u]/(u^{2k}-1)$. Finite Fields Appl. 46, 217-234 (2017)

%[17]
%\bibitem{src18}Gao, J., Gao, Y., Fu, F.: Quantum codes from cyclic codes over the ring $\mathbb{F}_{q}+v_{1}\mathbb{F}_{q}+...+v_{r}\mathbb{F}_{q}$. Appl. Algebra Engrg. Comm. Comput. 30, 161-174 (2019)

%[18]
%\bibitem{GF17&19}Gulliver, T., Venakaiah, V.: Construction of quasi-twisted codes and enumeration of defining polynomials. J. Algebra Comb. Discrete Struct. Appl. 7, 1-18 (2019)

%[19]
%\bibitem{src6}Bag, T., Dinh, H. Q., Upadhyay, A. K., Yamaka, W.: New non-binary quantum codes from cyclic codes over product rings. IEEE Communications Letters. 24, 486-490 (2019)

%[20]
%\bibitem{src32} Qian, J., Zhang, L.: Nonbinary quantum codes derived from repeated-root cyclic codes, Modern Phys. Lett. B. 27, 1350053 (2013)

%\bibitem{seq} Hou, X., Lopez-Permouth, S., Parra-Avila, B.: Rational power series, sequential codes and periodicity of sequences. J. Pure Appl. Algebra. 213, 1157-1169 (2009)

%\bibitem{1972} Peterson, W. W., Weldon, E. J.: Error Correcting Codes. MIT Press (1972)

%[23]
%\bibitem{src14} Bag, T., Bandi, R., Chinnakum, W., Dinh, H., Upadhyay, A.: On the structure of cyclic codes over $F_{q}RS$ and applications in quantum and LCD codes constructions. IEEE Access. 8, 18902-18914 (2020)

%[24]
%\bibitem{src30}Özen, M., Özzaim, T., İnce, H.: Skew quasi cyclic codes over $\mathbb{F}_{q}+v\mathbb{F}_{q}$. J. Algebra Appl. 18, 1950077 (2018)

%[25]
%\bibitem{src29} Fu, F., Gao, J., Ma F.: New non-binary quantum codes from constacyclic codes over $\mathbb{F}_q[u,v]/\langle u^{2}-1, v^{2}-v, uv-vu\rangle$. Adv. Math. Commun. 13, 421-434 (2019)

%[26]
%\bibitem{src5}Ashraf, M., Bag, T., Mohammad, G., Upadhyay, A.: Quantum codes from cyclic codes over the ring $\mathbb{F}_p[u] / \langle u^3-u \rangle$. Asian-Eur. J. Math. 12, 2050008 (2020)

%[27]
%\bibitem{src26}Koroglu, M., Siap, I.: Quantum Codes From A Class of Constacyclic Codes over Group Algebras. Malays. J. of Math. Sci. 11, 289-301 (2017)

%[28]
%\bibitem{src19} Fu, F., Gao, J., Ma, F.: Constacyclic codes over the ring ${\mathbb {F}}_q+v{\mathbb {F}}_q+v^{2}{\mathbb {F}}_q$ and their applications of constructing new non-binary quantum codes. Quantum Inf. Process. 17, 122 (2018)

%[29]
%\bibitem{src13} Bag, T., Dinh, H., Upadhyay, A., Ashraf, M., Mohammad, G., Chinnakum, W.: New quantum codes from a class of constacyclic codes over Finite Commutative Rings. J. Algebra Appl. 19, 2150003 (2020)



\bibitem{smallkeys} Heyse, Stefan, von Maurich, Ingo
and G{\"u}neysu, Tim", Smaller Keys for Code-Based Cryptography: QC-MDPC McEliece Implementations on Embedded Devices, in "Cryptographic Hardware and Embedded Systems - CHES 2013", editors"Bertoni, Guido and Coron, Jean-S{\'e}bastien", Springer Berlin Heidelberg, Berlin, Heidelberg, 2013, pp273--292.

\bibitem{bible} MacWilliams, F. J., Sloane N. J.: The theory of error correcting codes. Elsevier; 1977.

\bibitem{constX} Sloane, N., Reddy, S., and  Chen, C-L.: New binary codes, IEEE Transactions on Information Theory 18(4), 503-510, (1972)






\end{thebibliography}
\end{document}